\documentclass[prd,showpacs,preprintnumbers,twocolumn,amsmath,nofootinbib,amssymb]{revtex4}
\usepackage{graphicx,color,dcolumn,booktabs,bm}
\usepackage{longtable,lscape}
\usepackage{txfonts}
\usepackage{overpic}
\usepackage{amssymb}
\usepackage{epstopdf}
\usepackage{indentfirst}
\usepackage{feynmf}   
\usepackage{slashed}  
\usepackage{cases}
\usepackage{color}
\usepackage{float}
\usepackage{multirow}
\usepackage{graphicx,color,dcolumn,booktabs,bm}
\usepackage{epsfig,dsfont,amssymb,amsmath,amsfonts,amsbsy,mathrsfs}

\graphicspath{{Figures/}} %

\usepackage{hyperref}
\hypersetup{colorlinks,
  citecolor=blue,anchorcolor=red,menucolor=red, linkcolor=red,filecolor=red,runcolor=red,urlcolor=blue,frenchlinks=red
}

\makeatletter
\@addtoreset{equation}{section}
\makeatother

\allowdisplaybreaks

\begin{document}

\title{Prediction of triple-charm molecular pentaquarks}

\author{Rui Chen$^{1,2,3}$}
\email{chenr15@lzu.edu.cn}
\author{Atsushi Hosaka$^{3}$}
\email{hosaka@rcnp.osaka-u.ac.jp}
\author{Xiang Liu$^{1,2}$}
\email{xiangliu@lzu.edu.cn}
\affiliation{
$^1$School of Physical Science and Technology, Lanzhou University, Lanzhou 730000, China\\
$^2$Research Center for Hadron and CSR Physics, Lanzhou University
and Institute of Modern Physics of CAS, Lanzhou 730000, China\\
$^3$Research Center for Nuclear Physics (RCNP), Osaka University, Ibaraki, Osaka 567-0047, Japan
}

\begin{abstract}
  In a one-boson-exchange model, we study molecular states of double-charm baryon ($\Xi_{cc}(3621)$) and a charmed meson ($D$ and $D^*$). Our model indicates that there exist two possible triple-charm molecular pentaquarks, a $\Xi_{cc}D$ state with $I(J^P)=0(1/2^-)$ and a $\Xi_{cc}D^*$ state with $I(J^P)=0(3/2^-)$. In addition, we also extend our formula to explore $\Xi_{cc}\bar{B}^{(*)}$, $\Xi_{cc}\bar{D}^{(*)}$, and $\Xi_{cc}B^{(*)}$ systems, and find more possible heavy flavor molecular pentaquarks, a $\Xi_{cc}\bar{B}$ state with $I(J^P)=0(1/2^-)$, a $\Xi_{cc}\bar{B}^*$ state with $I(J^P)=0(3/2^-)$, and  $\Xi_{cc}\bar{D}^*/\Xi_{cc}B^*$ states with $I(J^P)=0(1/2^-)$. Experimental search for these predicted triple-charm molecular pentaquarks is encouraged.
\end{abstract}


\maketitle

\section{introduction}\label{sec1}

Recently, the LHCb Collaboration reported double-charm baryon $\Xi_{cc}^{++}$ in the $\Lambda_c^+K^-\pi^+\pi^-$ invariant mass spectrum \cite{Aaij:2017ueg}, which has the mass $3621.40\pm0.72\pm0.27\pm0.14$ MeV. Obviously, the observation of $\Xi_{cc}^{++}$ plays a crucial role to establish complete baryon family. The announced $\Xi_{cc}^{++}(3621)$ by LHCb may provide input to carry out the research issues around double-charm baryon.

\begin{figure}[!htbp]
\center
\includegraphics[width=3.4in]{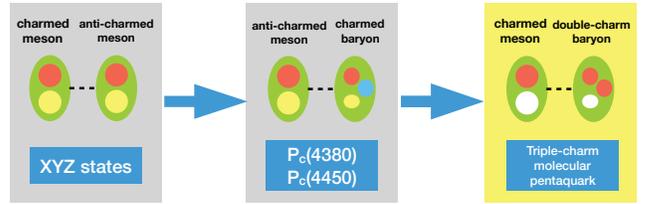}
\caption{Evolution of interaction of hadrons and the corresponding connections with charmoium-like $XYZ$ states, $P_c(4380)/P_{c}(4450)$ and triple-charm molecular pentaquark.}\label{fig:1}
\end{figure}

In the past decade, hadronic molecular states were extensively explored with the discovery of charmonium-like $XYZ$ states and $P_c(4380)$ and $P_c(4450)$.  The observed charmonium-like $XYZ$ states have stimulated the studies of the interaction between charmed meson and anti-charmed meson, while these reported $P_c$ states make that this study can be extended to interaction between charmed meson and charmed baryon. The observation of double-charm baryon $\Xi_{cc}^{++}(3621)$ inspires our interest in exploring double-charm baryon interacting with a charmed meson.
As shown in Fig. \ref{fig:1}, investigating the interaction between charmed meson and double-charm baryon is a natural extension of former studies involved in charmed meson interacting with charmed baryon. By this study, we would like to study whether or not there exist possible triple-charm molecular pentaquarks, which can arouse experimentalist's interest in searching such new type of exotic hadronic molecular states after the LHCb experimental observation of double-charm baryon.

In this work, we mainly focus on the interaction between S-wave double-charm baryon $\Xi_{cc}(3621)$ and S-wave charmed meson ($D$ and $D^*$) and study possible triple-charm molecular pentaquarks. This dynamical interaction can be quantitatively described by one-boson-exchange (OBE) model, which was often adopted to investigate the hadronic molecular states \cite{Liu:2008fh,Thomas:2008ja,Lee:2009hy,Liu:2009ei,Sun:2011uh,Chen:2015loa}. By OBE model, the effective potential of the interaction between S-wave double-charm baryon $\Xi_{cc}(3621)$ and S-wave charmed meson ($D$ and $D^*$) can be extracted, by which we may search for the corresponding molecular states. This information can encourage experimental studies of searching for triple-charm molecular pentaquarks.

An outline of this paper is as follows. In Sec. \ref{sec2}, we present the detailed derivation of effective potential related to the interaction between an S-wave $\Xi_{cc}$ baryon and an S-wave charmed meson $D/D^*$, and the corresponding numerical result are presented in \ref{sec3}.
Finally, we will give a summary in Sec. \ref{sec4}.

\section{The details of the obtained effective potentials}\label{sec2}

\subsection{OBE potentials}

\begin{figure}[!htbp]
\center
\includegraphics[width=2in]{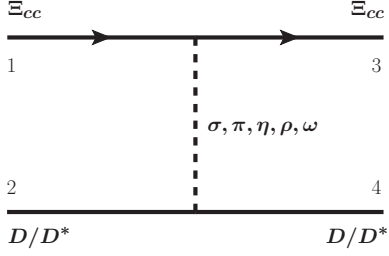}\\
\caption{ The typical diagram for the $\Xi_{cc}D^{(*)}$ systems with the one-boson-exchange model.}\label{obe}
\end{figure}

Our one boson exchange model consists of light meson exchanges $\pi$, $\eta$, $\sigma$, $\rho$, and $\omega$ as shown in Fig. \ref{obe}. The structure of the interaction Lagrangian is determined by the quantum numbers, and in particular, we remind that $\Xi_{cc}(3621)$ has $J^P=1/2^+$ \cite{Chen:2017sbg}. The relevant Lagrangians for three-meson vertices are given by
\begin{eqnarray}
\mathcal{L}_{D^{(*)}D^{(*)}\sigma} &=& -2g_s D_bD_b^{\dag}\sigma+2g_sD_b^{*}\cdot D_b^{*\dag}\sigma,\label{lag1}\\
\mathcal{L}_{D^{*}D^{*}\mathbb{P}} &=& -i\frac{2g}{f_{\pi}}v^{\alpha}\varepsilon_{\alpha\mu\nu\lambda}
D_b^{*\mu}D_a^{*\lambda\dag}\partial^{\nu}\mathbb{P}_{ba},\\
\mathcal{L}_{D^{(*)}D^{(*)}\mathbb{V}} &=& -\sqrt{2}\beta g_V D_b D_a^{\dag} v\cdot\mathbb{V}_{ba}
    +\sqrt{2}\beta g_V D_b^*\cdot D_a^{*\dag}v\cdot\mathbb{V}_{ba}\nonumber\\
   &&-i2\sqrt{2}\lambda g_V D_b^{*\mu}D_a^{*\nu\dag}
   \left(\partial_{\mu}\mathbb{V}_{\nu}-\partial_{\nu}\mathbb{V}_{\mu}\right)_{ba},\label{lag2}
\end{eqnarray}
which can be constructed by considering the requirement of the heavy quark symmetry and chiral symmetry \cite{Yan:1992gz,Wise:1992hn,Burdman:1992gh,Casalbuoni:1996pg,Falk:1992cx}.

In Refs. \cite{Meng:2017fwb,Meng:2017udf}, the effective Lagrangians for the coupling of S-wave double-charm baryons with light mesons are constructed as
\begin{eqnarray}
\mathcal{L}_{\Xi_{cc}\Xi_{cc}\sigma} &=& g_{\sigma}\bar{\Xi}_{cc}\sigma\Xi_{cc},\label{lag3}\\
\mathcal{L}_{\Xi_{cc}\Xi_{cc}\mathbb{P}} &=& g_{\pi}\bar{\Xi}_{cc}i\gamma_5\mathbb{P}\Xi_{cc},\\
\mathcal{L}_{\Xi_{cc}\Xi_{cc}\mathbb{V}} &=& h_{v}\bar{\Xi}_{cc}\gamma_{\mu}\mathbb{V}^{\mu}\Xi_{cc}
          +\frac{f_v}{2M_{\Xi_{cc}}}\bar{\Xi}_{cc}\sigma_{\mu\nu}\partial^{\mu}\mathbb{V}^{\nu}\Xi_{cc}.\label{lag4}
\end{eqnarray}
Here, $\mathbb{P}$ and $\mathbb{V}$ in above Lagrangians correspond to the pseudoscalar and vector matrixes, respectively, i.e.,
\begin{eqnarray*}
\mathbb{P} = \left(\begin{array}{cc}
\frac{\pi^0}{\sqrt{2}}+\frac{\eta}{\sqrt{6}} &\pi^+\\
\pi^- &-\frac{\pi^0}{\sqrt{2}}+\frac{\eta}{\sqrt{6}}
\end{array}\right),\,
\mathbb{V} = \left(\begin{array}{cc}
\frac{\rho^0}{\sqrt{2}}+\frac{\omega}{\sqrt{2}}  &\rho^+\\
\rho^- &-\frac{\rho^0}{\sqrt{2}}+\frac{\omega}{\sqrt{2}}
\end{array}\right).
\end{eqnarray*}

With above effective Lagrangians, one further writes out the scattering amplitudes for the processes $\Xi_{cc}D\to\Xi_{cc}D$ and $\Xi_{cc}D^*\to\Xi_{cc}D^*$ showed in Fig. \ref{obe}. And then, the effective potential in the momentum space can be related to the scattering amplitude by the Breit approximation, where the general relation is
\begin{eqnarray}\label{breit}
\mathcal{V}^{h_1h_2\to h_3h_4}(\bm{q}) &=&
          -\frac{\mathcal{M}(h_1h_2\to h_3h_4)}
          {\sqrt{\prod_i2M_i\prod_f2M_f}},
\end{eqnarray}
where $M_i$ and $M_f$ denote the masses of the initial states ($h_1$, $h_2$) and final states ($h_3$, $h_4$), respectively. $\mathcal{M}(h_1h_2\to h_3h_4)$ is the scattering amplitude for the $h_1h_2\to h_3h_4$ process.
Finally, the effective potential in the coordinate space $\mathcal{V}(\bm{r})$ can be extracted by performing Fourier transformation, i.e.,
\begin{eqnarray}
\mathcal{V}^{h_1h_2\to h_3h_4}(\bm{r}) =
          \int\frac{d^3\bm{q}}{(2\pi)^3}e^{i\bm{q}\cdot\bm{r}}
          \mathcal{V}^{h_1h_2\to h_3h_4}(\bm{q})\mathcal{F}^2(q^2,m_E^2).\nonumber\\\label{vr}
\end{eqnarray}
In Eq. (\ref{vr}), a monopole form factor $\mathcal{F}(q^2,m_E^2)=(\Lambda^2-m_E^2)/(\Lambda^2-q^2)$ is introduced at every interaction vertex. It can reflect finite size effect of the hadrons involved in these interactions. Here, $\Lambda$, $m_E$ and $q$ are the cutoff, mass and four-momentum of the exchanged meson, respectively.

In the following, we collect the expressions of the total effective potentials for these discussed $\Xi_{cc}D$ and $\Xi_{cc}D^*$ systems, which include
\begin{widetext}
\begin{eqnarray}
V_{\Xi_{cc}D}^I(r) &=& g_sg_{\sigma}Y(\Lambda,m_{\sigma},r)
      -\left[\frac{\beta g_V h_{v}}{\sqrt{2}}+\frac{\beta g_V f_v}{4\sqrt{2}m_{\Xi_{cc}}^2}\mathcal{O}(r)
        \right]
      \left[\mathcal{G}(I)Y(\Lambda,m_{\rho},r)+\frac{1}{2}Y(\Lambda,m_{\omega},r)\right],\label{D}\\
V_{\Xi_{cc}D^*}^I(r) &=& g_sg_{\sigma}Y(\Lambda,m_{\sigma},r)
      +\frac{gg_{\pi}}{6f_{\pi}m_{\Xi_{cc}}}\left[i\bm{\sigma}\cdot\left(\bm{\epsilon_2}\times\bm{\epsilon_4^{\dag}}\right)
          \mathcal{O}(r)+S(\hat{r},\bm{\sigma},i\bm{\epsilon_2}\times\bm{\epsilon_4^{\dag}})\mathcal{P}(r)\right]
      \left[\mathcal{G}(I)Y(\Lambda,m_{\pi},r)+\frac{1}{6}Y(\Lambda,m_{\eta},r)\right]\nonumber\\
      &&+\left[-\frac{\beta g_Vh_v}{\sqrt{2}}
        -\frac{\beta g_V f_v}{4\sqrt{2}m_{\Xi_{cc}}^2}\mathcal{O}(r)
        -\frac{\sqrt{2}\lambda g_V(h_v+f_v)}{3m_{\Xi_{cc}}}i\bm{\sigma}\cdot\left(\bm{\epsilon_2}\times\bm{\epsilon_4^{\dag}}\right)
        \mathcal{O}(r)+\frac{\lambda g_V(h_v+f_v)}{3\sqrt{2}m_{\Xi_{cc}}}
        S(\hat{r},\bm{\sigma},i\bm{\epsilon_2}\times\bm{\epsilon_4^{\dag}})\mathcal{P}(r)\right.\nonumber\\
        &&\left.+\frac{\beta g_V f_v}{2\sqrt{2}m_{\Xi_{cc}}}\bm{\sigma}\cdot\bm{L}
        \bm{\epsilon_2}\cdot\bm{\epsilon_4^{\dag}}\mathcal{Q}(r)+\frac{\sqrt{2}\lambda g_V h_v}{m_{\Xi_{cc}}}i\left(\bm{\epsilon_2}\times\bm{\epsilon_4^{\dag}}\right)\cdot\bm{L}\mathcal{Q}(r)\right]
        \left[\mathcal{G}(I)Y(\Lambda,m_{\rho},r)+\frac{1}{2}Y(\Lambda,m_{\omega},r)\right].\label{DS}
\end{eqnarray}
\end{widetext}
Here, $I$ stands for the isospin for the $\Xi_{cc}D^{(*)}$ systems. $\mathcal{G}(I)$ is isospin factor, which is taken as $3/2$ for isoscalar systems, and $-1/2$ for isovector systems. The flavor wave functions $|I,I_3\rangle$ involved in this work are
\begin{eqnarray}
|1,1\rangle &=& |\Xi_{cc}^{++}D^{(*)+}\rangle,\\
|1,0\rangle &=& \frac{1}{\sqrt{2}}\left(|\Xi_{cc}^{++}D^{(*)0}-\Xi_{cc}^+D^{(*)+}\rangle\right),\\
|1,-1\rangle &=& |\Xi_{cc}^+D^{(*)0}\rangle,\\
|0,0\rangle &=& \frac{1}{\sqrt{2}}\left(|\Xi_{cc}^{++}D^{(*)0}+\Xi_{cc}^+D^{(*)+}\rangle\right).
\end{eqnarray}

In Eqs. (\ref{D})-(\ref{DS}), we define these operators $\mathcal{O}(r)$, $\mathcal{P}(r)$, $\mathcal{Q}(r)$, $S(\hat{r},\bm{a},\bm{b})$ and function $Y(\Lambda, m,r)$, which can be further expressed as
\begin{eqnarray*}
&&\mathcal{O}(r) = \frac{1}{r^2}\frac{\partial}{\partial r}r^2\frac{\partial}{\partial r},\quad
\mathcal{P}(r) = r\frac{\partial}{\partial r}\frac{1}{r}\frac{\partial}{\partial r},\quad \mathcal{Q}(r)=\frac{1}{r}\frac{\partial}{\partial r}.\\
&&Y(\Lambda,m,r) = \frac{1}{4\pi r}(e^{-mr}-e^{-\Lambda r})-\frac{\Lambda^2-m^2}{8\pi \Lambda}e^{-\Lambda r},\\
&&S(\hat{r},\bm{a},\bm{b})=3(\hat{r}\cdot\bm{a})(\hat{r}\cdot\bm{b})-\bm{a}\cdot\bm{b},
\end{eqnarray*}
respectively.

The study on the deuteron indicates that the S-D mixing related to tensor force is very important \cite{Tornqvist:1993ng,Tornqvist:1993vu}. In this work, the S-D mixing effect will be also taken into account. Thus, the spin-orbit wave functions $|{}^{2S+1}L_J\rangle$ for the $\Xi_{cc}D^{(*)}$ systems involved in S-D mixing effect are
\begin{eqnarray*}
\left.\begin{array}{rl}
\Xi_{cc}D(1/2^-):     &\quad|{}^2S_{1/2}\rangle,\\
\Xi_{cc}D^*(1/2^-):   &\quad|{}^2S_{1/2}\rangle, \quad |{}^4D_{1/2}\rangle,\\
\Xi_{cc}D^*(3/2^-):   &\quad|{}^4S_{3/2}\rangle, \quad |{}^2D_{3/2}\rangle, \quad |{}^4D_{3/2}\rangle.
\end{array}\right.
\end{eqnarray*}
In addition, the general expressions of the spin-orbit wave functions for the $\Xi_{cc}D^{(*)}$ systems are constructed as
\begin{eqnarray}
\Xi_{cc}D:\, \left|{}^{2S+1}L_{J}\right\rangle &=&
          \chi_{\frac{1}{2}m_S}|Y_{L,m_L}\rangle,\nonumber\\
\Xi_{cc}D^*: \left|{}^{2S+1}L_{J}\right\rangle &=&
\sum_{m,m'}^{m_S,m_L}C^{S,m_S}_{\frac{1}{2}m,1m'}C^{J,M}_{Sm_S,Lm_L}
          \chi_{\frac{1}{2}m}\epsilon^{m'}|Y_{L,m_L}\rangle.\nonumber
\end{eqnarray}
Here, $C^{J,M}_{Sm_S,Lm_L}$ and $C^{S,m_S}_{\frac{1}{2}m,1m'}$ are the Clebsch-Gordan coefficients. $\epsilon$, $\chi_{\frac{1}{2}m}$ and $Y_{L,m_L}$ are defined as the polarization vector, spin wave function and spherical harmonics function, respectively.

For the spin-spin, spin-orbit and tensor force operators in Eq. (\ref{DS}), they should be sandwiched by the above spin-orbit wave functions, like $\langle\Xi_{cc}D^* ({}^2S_{1/2})|i\bm{\sigma}\cdot(\bm{\epsilon_2}\times\bm{\epsilon_4^{\dag}})|\Xi_{cc}D^* ({}^2S_{1/2})\rangle$. The obtained relevant numerical matrices are summarized in Table \ref{operator}.

\renewcommand\tabcolsep{0.7cm}
\renewcommand{\arraystretch}{1.7}
\begin{table}[!htbp]
\caption{Matrix elements for $\langle f|\mathcal{A}|i\rangle$ for the operators $\mathcal{A}$.}\label{operator}
{\begin{tabular}{cc}
\toprule[1pt]
$\left\langle i\bm{\sigma}\cdot(\bm{\epsilon_2}\times\bm{\epsilon_4^{\dag}})\right\rangle_{J=1/2}$
      &$\left\langle S(\hat{r},\bm{\sigma},i\bm{\epsilon_2}\times\bm{\epsilon_4^{\dag}})\right\rangle_{J=1/2}$\\\hline
$\left(\begin{array}{cc} -2    &0\\    0    &1\end{array}\right)$
      &$\left(\begin{array}{cc} 0    &-\sqrt{2}\\    -\sqrt{2}    &-2\end{array}\right)$\\\hline
$\left\langle \bm{\sigma}\cdot\bm{L}\bm{\epsilon_2}\cdot\bm{\epsilon_4^{\dag}}\right\rangle_{J=1/2}$
      &$\left\langle i(\bm{\epsilon_2}\times\bm{\epsilon_4^{\dag}})\cdot\bm{L}\right\rangle_{J=1/2}$\\\hline
$\left(\begin{array}{cc} 0    &0\\    0    &-3\end{array}\right)$
      &$\left(\begin{array}{cc} 0    &0\\    0    &-3\end{array}\right)$\\\hline
$\left\langle i\bm{\sigma}\cdot(\bm{\epsilon_2}\times\bm{\epsilon_4^{\dag}})\right\rangle_{J=3/2}$
      &$\left\langle S(\hat{r},\bm{\sigma},i\bm{\epsilon_2}\times\bm{\epsilon_4^{\dag}})\right\rangle_{J=3/2}$\\\hline
$\left(\begin{array}{ccc} 1   &0   &0\\   0   &-2   &0\\    0    &0   &1\end{array}\right)$
               &$\left(\begin{array}{ccc} 0   &1   &2\\   1   &0   &-1\\    2    &-1   &0\end{array}\right)$\\\hline
$\left\langle \bm{\sigma}\cdot\bm{L}\bm{\epsilon_2}\cdot\bm{\epsilon_4^{\dag}}\right\rangle_{J=3/2}$
       &$\left\langle i(\bm{\epsilon_2}\times\bm{\epsilon_4^{\dag}})\cdot\bm{L}\right\rangle_{J=3/2}$\\\hline
$\left(\begin{array}{ccc} 0   &0   &0\\   0   &1   &-2\\    0    &-2   &-2\end{array}\right)$
       &$\left(\begin{array}{ccc} 0   &0   &0\\   0   &-2   &1\\    0    &1   &-2\end{array}\right)$
\\
\bottomrule[1pt]
\end{tabular}}
\end{table}

\subsection{Parameters}

In this work, the parameters are the coupling constants, cutoff $\Lambda$ and masses of the particles.
Let us first look at the coupling constants in Eqs. (\ref{lag1})-(\ref{lag4}). For the charmed meson sector, there are four coupling constants ($g_s$, $g$, $\beta$, $\lambda$), a pion decay constant $f_{\pi}=132$ MeV and $g_V=m_{\rho}/f_{\pi}=5.8$. Due to spontaneously broken chiral symmetry \cite{Bardeen:2003kt}, $g_s$ is related to the coupling constant $\tilde g$ for the process $D(0^+)\to D(0^-)+\pi$, $g_s=\tilde g/2\sqrt{6}$, where $\tilde g$ is taken as 3.73 in Ref. \cite{Falk:1992cx}. In heavy quark symmetry, the coupling constant for $D^*D^*\pi$ interaction is taken the same value of $D^*D\pi$ coupling, $g=0.59$, which is extracted from the decay width of $D^{*+}$ \cite{Isola:2003fh}. According to vector meson dominance \cite{Isola:2003fh}, $\beta$ is fixed as $\beta=$0.9. Through a comparison of the form factor between the theoretical calculation from light cone sum rule and lattice QCD, the numerical value for $\lambda$ is determined as 0.56 GeV$^{-1}$ \cite{Isola:2003fh}.

With the help of the quark model, coupling constants for the interaction of double-charm baryons and light meson are derived from nucleon-nucleon interaction, i.e.,
\begin{eqnarray}
\mathcal{L}_N &=& g_{\sigma NN}\bar{N}\sigma N+\sqrt{2}g_{\pi NN} \bar{N}i\gamma_5\mathbb{P}N\nonumber\\
                 &&+\sqrt{2}g_{\rho NN}\bar{N}\gamma_{\mu}\mathbb{V}^{\mu}N+\frac{f_{\rho NN}}{\sqrt{2}m_N}\bar{N}\sigma_{\mu\nu}\partial^{\mu}\mathbb{V}^{\nu}N.\label{lag5}
\end{eqnarray}
We can obtain several relations of coupling constants, i.e.,
\begin{eqnarray}
&&g_{\sigma}=\frac{1}{3}g_{\sigma NN},\quad g_{\pi}=-\frac{\sqrt{2}}{5}\frac{m_{\Xi_{cc}}}{m_N}g_{\pi NN},\quad
h_{v}=\sqrt{2}g_{\rho NN},\nonumber\\
&&h_{v}+f_{v}=-\frac{\sqrt{2}}{5}\frac{m_{\Xi_{cc}}}{m_N}(g_{\rho NN}+f_{\rho NN}).
\end{eqnarray}
Coupling constants for the nucleon-nucleon interaction are given in Refs. \cite{Machleidt:2000ge,Machleidt:1987hj,Cao:2010km}.
In Table \ref{coupling}, we collect numerical values for all the coupling constants.
\renewcommand\tabcolsep{0.3cm}
\renewcommand{\arraystretch}{2}
\begin{table}[!htbp]
  \caption{A summary of coupling constants. The signs for the coupling constants are fixed by the quark model.}\label{coupling}
  \begin{tabular}{llll}\toprule[1pt]
  $\sigma$    &$\pi/\eta$     &$\rho/\omega$\\\hline
  $g_s=0.76$  &$g/{f_{\pi}}=4.47$        &$\beta g_V=5.22$     &$\lambda g_V=3.25$\\\hline
  $\frac{g_{\sigma NN}^2}{4\pi}=5.69$         &$\frac{g_{\pi NN}^2}{4\pi}=13.60$         &$\frac{g_{\rho NN}^2}{4\pi}=0.84$          &$\frac{f_{\rho NN}}{g_{\rho NN}}=6.10$    \\\hline
  $g_{\sigma}=-2.82$      &$g_{\pi}=14.26$     &$h_{v}=4.65$    &$f_{v}=-28.11$   \\
  \bottomrule[1pt]
  \end{tabular}
\end{table}

Since the discussed hadrons are not point-like particles, we should introduce the monopole form factor in each interactive vertex. 
For cutoff $\Lambda$, we recall a cutoff relation on charge root-mean-square radius of the interactive hadron, $\Lambda_1/\Lambda_2=\sqrt{\langle r_2^2 \rangle/\langle r_1^2 \rangle}$, which is based on
\begin{eqnarray}
\langle r^2 \rangle &\equiv& \left.\frac{-6}{\mathcal{F}(q^2,m_E^2)}\frac{\partial\mathcal{F}(q^2,m_E^2)}{\partial q^2}\right|_{q^2\to 0} \approx \frac{6}{\Lambda^2}.
\end{eqnarray}
In Ref. \cite{Yasui:2009bz}, cutoff $\Lambda$ for the charm and bottom mesons are estimated by
\begin{eqnarray}
\Lambda_D= 1.35\Lambda_N, \quad \Lambda_B=1.29 \Lambda_N, \label{cutoff}
\end{eqnarray}
where $\Lambda_N$, $\Lambda_D$ and $\Lambda_B$ correspond to the cutoffs for nucleon, $D$ and $B$ mesons, respectively. To fix the cutoff of nucleon $\Lambda_N$, we would like to adopt the OBE model to reproduce the binding energy of deuteron. According to the effective Lagrangian in Eq. (\ref{lag5}), the OBE effective potential for nucleon-nucleon system with $I(J^P)=0(1^+)$ is
\begin{eqnarray}
V(r) &=& -g_{\sigma NN}^2\bm{\sigma}_1\cdot\bm{\sigma}_2Y(\Lambda,m_{\sigma},r)\nonumber\\
    &&+\frac{g_{\pi NN}^2}{4m_N^2}\left[\bm{\sigma}_1\cdot\bm{\sigma}_2\mathcal{O}(r)+S(\hat{r},\bm{\sigma}_1,\bm{\sigma}_2)\mathcal{P}(r)\right]\nonumber\\
    &&\times\left[-Y(\Lambda,m_{\pi},r)+\frac{1}{9}Y(\Lambda,m_{\eta},r)\right]\nonumber\\
    &&+\left[g_{\rho NN}^2\bm{\sigma}_1\cdot\bm{\sigma}_2+\frac{f_{\rho NN}^2}{2m_N^2}\bm{\sigma}_1\cdot\bm{\sigma}_2\mathcal{O}(r)\right.\nonumber\\
    &&\left.+\frac{g_{\rho NN}^2}{2m_N^2}(\bm{\sigma}_1+\bm{\sigma}_2)\cdot\bm{L}\mathcal{Q}(r)\right.\nonumber\\
    &&\left.-\frac{g_{\rho NN}^2+2g_{\rho NN}f_{\rho NN}-2f_{\rho NN}^2}{12m_N^2}\bm{\sigma}_1\cdot\bm{\sigma}_2\mathcal{O}(r)\right.\nonumber\\
    &&\left.-\frac{g_{\rho NN}^2-g_{\rho NN}f_{\rho NN}+f_{\rho NN}^2}{12m_N^2}S(\hat{r},\bm{\sigma}_1,\bm{\sigma}_2)\mathcal{P}(r)\right]\nonumber\\
    &&\times\left[-3Y(\Lambda,m_{\rho},r)+Y(\Lambda,m_{\omega},r)\right].\label{nn}
\end{eqnarray}
Under considered the S-D mixing effect, operators for spin-spin, spin-orbit and tensor force interactions in Eq. (\ref{nn}) are replaced by
\begin{eqnarray}
&&\langle\bm{\sigma}_1\cdot\bm{\sigma}_2\rangle            \mapsto          \left(\begin{array}{cc}1   &0\\    0   &1\end{array}\right),\quad
\langle S(\hat{r},\bm{\sigma}_1,\bm{\sigma}_2)\rangle    \mapsto          \left(\begin{array}{cc}0   &\sqrt{8}\\    \sqrt{8}   &-2\end{array}\right),\nonumber\\
&&\langle(\bm{\sigma}_1+\bm{\sigma}_2)\cdot\bm{L}\rangle   \mapsto          \left(\begin{array}{cc}0   &0\\    0   &-6\end{array}\right),
\end{eqnarray}
respectively.

By solving the coupled channel Schr\"{o}dinger equation, we can reproduce deuteron binding energy $E=-2.23$ MeV and obtain the corresponding root-mean-square radius $r_{RMS}=3.74$ fm, when $\Lambda_N$ is taken as $0.862$ GeV. Thus, by the cutoff relations in Eq. (\ref{cutoff}), $\Lambda_D$ is estimated as $1.164$ GeV.

Additionally, the masses and spin-parity quantum numbers for all involved hadrons are listed in Table \ref{mass}.

\renewcommand\tabcolsep{0.06cm}
\renewcommand{\arraystretch}{1.8}
\begin{table}[!htbp]
\caption{Masses and spin-parity of the hadrons involved in our study \cite{Olive:2016xmw}.}\label{mass}
{\begin{tabular}{ccc|ccc}
\toprule[1pt]
{Hadrons}      &$I^G(J^P)$    &{Mass (MeV)}    &\multicolumn{1}{c}{Hadrons}       &$I(J^P)$        &{Mass (MeV)}\\\hline
$\sigma$       &$0^+(0^+)$    &600             &\multicolumn{1}{c}{$\pi$}         &$1^-(0^-)$      &137.27\\
$\eta$         &$0^+(0^-)$    &547.85          &\multicolumn{1}{c}{$\rho$}        &$1^+(1^-)$      &775.49\\
$\omega$       &$0^-(1^-)$    &782.65          &\multicolumn{1}{c}{$N$}           &$1/2(1/2^+)$    &938.27\\
$D$            &$1/2(0^-)$    &1867.21         &\multicolumn{1}{c}{$D^*$}         &$1/2(1^-)$      &2008.56\\
$B$            &$1/2(0^-)$    &5279.42         &\multicolumn{1}{c}{$B^*$}         &$1/2(1^-)$      &5325.2\\
$\Xi_{cc}$     &$1/2(1/2^+)$   &3621.4       \\
\bottomrule[1pt]
\end{tabular}}
\end{table}

\section{Numerical results}\label{sec3}

Before producing the numerical calculation, we firstly perform a qualitative analysis of the properties of the OBE effective potentials for the $\Xi_{cc}D^{(*)}$ systems. In general, the pseudoscalar meson $\pi$, scalar meson $\sigma$ and vector meson $\rho/\omega$ exchanges provide long-range, intermediate-range and short-range contribution for the interaction between S-wave $\Xi_{cc}$ and S-wave charmed meson, respectively.

For the $\Xi_{cc}D$ system, there do not exist the $\pi$ and $\eta$ exchanges since the $DD\pi/DD\eta$ interactions are forbidden by symmetry. The $\sigma$ and $\omega$ exchanges provide attractive forces \cite{Chen:2017vai}, while the effective potentials from the $\rho$ exchange are repulsive and attractive for the isovector and isoscalar systems, respectively, and have relation {$V_{isoscalar}=-3V_{isovector}$}.
Thus, the isoscalar $\Xi_{cc}D$ system can more easily form bound state than the isovector $\Xi_{cc}D$ system.

For the $\Xi_{cc}D^*$ system, besides considering the intermediate-range  and short-range interactions from the $\sigma/\rho/\omega$ exchanges, the $\pi$ exchange and S-D wave mixing effect are also included in our calculation. As showed in Eq. (\ref{DS}), the spin-spin interaction, spin-orbit interaction and tensor force are involved in the total OBE effective potentials for the $\Xi_{cc}D^*$ system. According to the experience of studying on deuteron \cite{Tornqvist:1993ng,Tornqvist:1993vu}, the long-range force from pion exchange, tensor force and S-D mixing effect are very important when proton and neutron form a loosely bound deuteron. Compared with the $\Xi_{cc}D$ system, more bound solutions for the $\Xi_{cc}D^*$ system can be found.

\begin{figure}[!htbp]
\center
\includegraphics[width=3.41in]{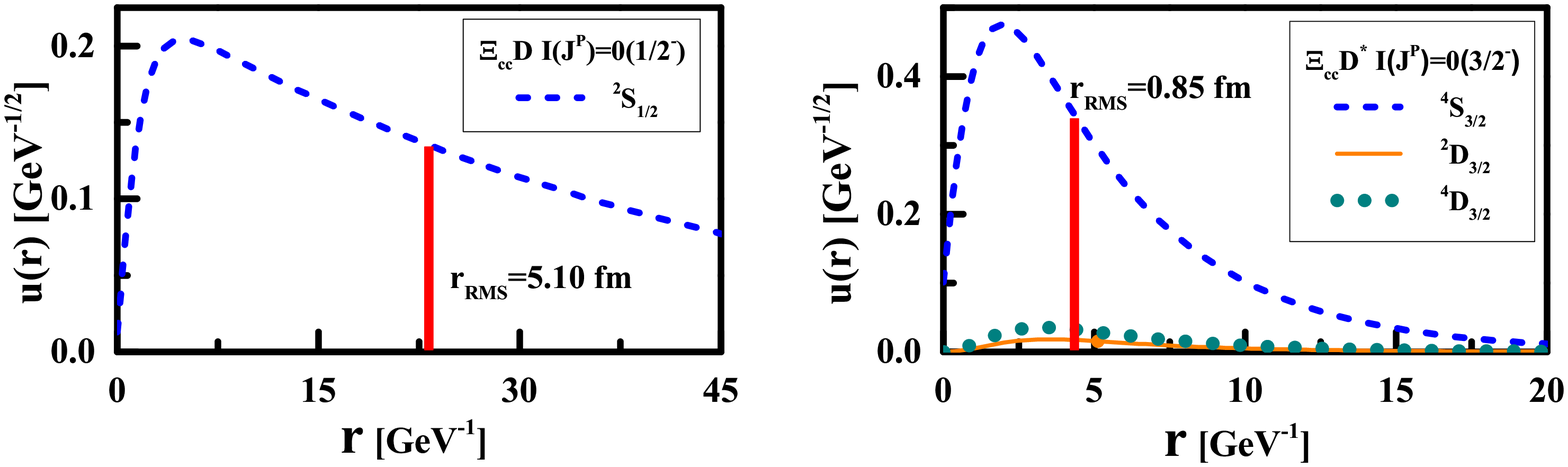}
\caption{The radial wave functions for the $\Xi_{cc}D$ state with $I(J^P)=0(1/2^-)$ and $\Xi_{cc}D^*$ state with $I(J^P)=0(3/2^-)$. The vertical lines locate in their root-mean-square radius $r_{RMS}$.}\label{wave}
\end{figure}

In the following, we present the concrete numerical results with the effective potentials in Eqs. (\ref{D})-(\ref{DS}) when cutoff $\Lambda_D$ is taken as 1.164 GeV. We find some interesting results:
\begin{enumerate}
  \item For the $\Xi_{cc}D$ system, an isoscalar state can be a molecular candidate, which has binding energy $E=-0.22$ MeV  and root-mean-square radius $r_{RMS}=5.10$ fm. In Fig. \ref{wave}, the $r$ dependence of its the radial wave function is shown, which reflects the $\Xi_{cc}D[0(1/2^-)]$ state to be a loosely bound triple-charm molecular state.

\item For the $\Xi_{cc}D^*$ system, there is only one good molecular candidate, the $\Xi_{cc}D^*$ state with $I(J^P)=0(3/2^-)$. Its binding energy is -18.71 MeV and root-mean-square radius is 0.85 fm. In comparison with the radial wave functions showed in Fig. \ref{wave}, we find that its dominant channel is $\Xi_{cc}D^*|{}^4S_{3/2}\rangle$.
\end{enumerate}

{Since the mass of $\Xi_{cc}D$ system is very close to the $\Xi_{cc}D^*$ system, and the pion exchange is also allowed for the process $\Xi_{cc}D^*\to\Xi_{cc}D$, we further consider the coupled channel effect for the $\Xi_{cc}D/\Xi_{cc}D^*$ system. With the same parameters input, we find that the binding energy for the $\Xi_{cc}D/\Xi_{cc}D^*$ system with $1/2^-$ is $E=-3.05$ MeV, and the dominant channel is $\Xi_{cc}D|{}^{2}S_{1/2}\rangle$ with probability around 99 percent. Thus, the coupled channel effect plays a rather minor role for the $\Xi_{cc}D/\Xi_{cc}D^*$ system.}

To summarize, two possible triple-charm molecular pentaquarks are predicted, which are the $\Xi_{cc}D$ state with $I(J^P)=0(1/2^-)$ and the $\Xi_{cc}D^*$ state with $I(J^P)=0(3/2^-)$. Our results also verify that the intermediate-range and short-range forces from $\sigma$, $\rho$, and $\omega$ exchanges are helpful in generating a bound state as suggested in Ref. \cite{Chen:2017vai}. The allowed strong decay models for these two possible triple-charm molecular pentaquarks are $\Omega_{ccc}\sigma$, $\Omega_{ccc}\omega$, $\Omega_{ccc}\pi\pi$. However, we need to indicate that the $\Omega_{ccc}$ baryon is still missing in experiment, which is big challenge to search for these predicted triple-charm molecular pentaquarks via these allowed decay channels.

\begin{figure}[!htbp]
\center
\includegraphics[width=3.2in]{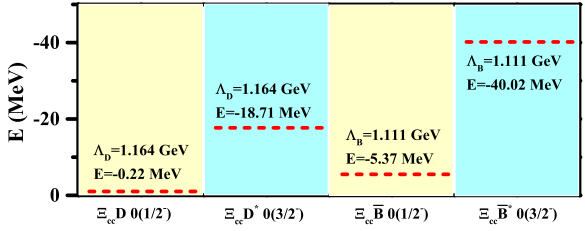}
\caption{The binding energies for the $\Xi_{cc}D^{(*)}$ and $\Xi_{cc}\bar{B}^{(*)}$ systems.}\label{energy}
\end{figure}

It is obvious that we may extend the obtained OBE effective potentials listed in Eqs. (\ref{D})-(\ref{DS}) to discuss the $\Xi_{cc}\bar{B}^{(*)}$ systems. Cutoff $\Lambda_B$ for the $\Xi_{cc}\bar{B}^{(*)}$ systems is taken as $1.11$ GeV which is determined by the relation in Eq. (\ref{cutoff}). Finally, we further predict two other possible molecular pentaquarks, the $\Xi_{cc}\bar{B}$ state with $I(J^P)=0(1/2^-)$, and the $\Xi_{cc}\bar{B}^*$ state with $I(J^P)=0(3/2^-)$. Their binding energies are given in Fig. \ref{energy}. Here, we want to specify two points: (1) Compared to the $\Xi_{cc}D/\Xi_{cc}\bar{B}$ states with $0(1/2^-)$, the $\Xi_{cc}D^*/\Xi_{cc}\bar{B}^*$ states with $0(3/2^-)$ are more stable, which implies the pion exchange and S-D mixing effect play important role in generating these discussed triple-charm molecular states.
(2) In general, hadrons with heavier masses are more easily bound, which can be reflected by our results, i.e., with the same dynamical interactions, we may find that the obtained binding energy of the $\Xi_{cc}\bar{B}^{(*)}$ system is larger than that of the $\Xi_{cc}D^{(*)}$ system.

In this paper, we further study the $\Xi_{cc}\bar{D}^{(*)}$ and $\Xi_{cc}B^{(*)}$ systems. According to the G-parity rule \cite{Klempt:2002ap}, the OBE effective potentials for the $\Xi_{cc}\bar{D}^{(*)}$ and $\Xi_{cc}B^{(*)}$ systems can related to the potentials for the $\Xi_{cc}{D}^{(*)}$ and $\Xi_{cc}\bar B^{(*)}$ systems in Eqs. (\ref{D})-(\ref{DS}). The interactions from the $\pi$ and $\omega$ exchanges are contrary to those for the $\Xi_{cc}{D}^{(*)}$ and $\Xi_{cc}\bar B^{(*)}$ systems.

In Table \ref{num}, we present bound solutions for the $\Xi_{cc}\bar D^{(*)}/\Xi_{cc}{B}^{(*)}$ systems.
For the $\Xi_{cc}\bar{D}/\Xi_{cc}B$ systems, the quantitative analysis does not support the existence of $\Xi_{cc}\bar D^{(*)}$ and $\Xi_{cc}{B}^{(*)}$ bound states, i.e., the total OBE effective potentials for the $\Xi_{cc}\bar{D}/\Xi_{cc}B$ systems are weakly attractive, since the $\omega$ exchange effective becomes repulsive. Indeed, our numerical results support such conjugation for the $\Xi_{cc}\bar{D}/\Xi_{cc}B$ systems.

\renewcommand\tabcolsep{0.15cm}
\renewcommand{\arraystretch}{1.6}
\begin{table}[!htbp]
  \caption{The bound solutions $[E, r_{RMS}]$ for the $\Xi_{cc}\bar{D}^{(*)}$ and $\Xi_{cc}B^{(*)}$ systems. Here, cutoff for $\Xi_{cc}\bar{D}^{(*)}$ and $\Xi_{cc}B^{(*)}$ are taken as 1.164 GeV, 1.111 GeV, respectively. The units for binding energy $E$ and root-mean-square radius $r_{RMS}$ are MeV, and fm, respectively. Notation $\times$ stands for no bound solutions.}\label{num}
  \begin{tabular}{c|cccc}\toprule[1pt]
  $\Xi_{cc}\bar{D}$       &$\Xi_{cc}\bar{D}^*$\\
  $0(1/2^-)/1(1/2^-)$           &$0(1/2^-)$       &$0(3/2^-)/1(1/2^-)/1(3/2^-)$\\\hline
  $\times$          &$[-0.67, 3.48]$     &$\times$    \\\midrule[1pt]
  $\Xi_{cc}B$       &$\Xi_{cc}B^*$\\
  $0(1/2^-)/1(1/2^-)$           &$0(1/2^-)$       &$0(3/2^-)/1(1/2^-)/1(3/2^-)$\\\hline
  $\times$          &$[-10.63, 0.91]$     &$\times$   \\
  \bottomrule[1pt]
  \end{tabular}
\end{table}

For the $\Xi_{cc}\bar{D}^*/\Xi_{cc}B^*$ systems, the properties of $\pi$ exchange and $\omega$ exchanges are very different from those for the $\Xi_{cc}{D}^*/\Xi_{cc}\bar B^*$ systems. Our result shows that the isoscalar $\Xi_{cc}\bar{D}^*$ and $\Xi_{cc}B^*$ systems with $J^P=1/2^-$ can be the possible molecular candidates. Moreover, the $\Xi_{cc}\bar{D}^*$ molecule with $I(J^P)=0(1/2^-)$ is loosely bound states.
Possible two-body strong decay channels for the $\Xi_{cc}\bar{D}^*/\Xi_{cc}B^*$ molecular states with $I(J^P)=0(1/2^-)$ include $\Lambda_c(1/2^{\pm})+\eta_c/B_c(0^-)$, $\Lambda_c(1/2^+)+J/\psi/B_c^*(1^-)$, respectively.


\section{summary}\label{sec4}

The observation of double-charm baryon \cite{Aaij:2017ueg} provides us a good opportunity to study possible triple-charm molecular pentaquarks composed by an S-wave double-charm baryon and an S-wave charmed meson. In Ref. \cite{Guo:2013xga}, F. K. Guo etc. once predicted the possible triple-heavy pentaquarks composed of a heavy meson and a doubly-heavy baryon by exploring the consequences of heavy flavour, heavy quark spin and heavy antiquark-diquark symmetries for hadronic molecules within an effective field theory framework. In this work, we explore the interaction between an S-wave double charm baryon $\Xi_{cc}$ and an S-wave charmed meson $D/D^*$ by adopting the OBE model, and further predict the existence of possible triple-charm molecular pentaquark, which include
the $\Xi_{cc}D$ state with $I(J^P)=0(1/2^-)$ and the $\Xi_{cc}D^*$ state with $I(J^P)=0(3/2^-)$.

In addition, the derived formula of the effective potentials for the $\Xi_{cc}D^{(*)}$ system can be extended to study the $\Xi_{cc}\bar{B}/\Xi_{cc}\bar{B}^*$ systems. Two possible molecular candidates are predicted, which are the $\Xi_{cc}\bar{B}$ state with $I(J^P)=0(1/2^-)$ and $\Xi_{cc}\bar{B}^*$ state with $I(J^P)=0(3/2^-)$. In addition, as a byproduct, we further extend our investigation to the $\Xi_{cc}\bar{D}^{(*)}/\Xi_{cc}B^{(*)}$ systems. Our results indicate that there can exist two molecular pentaquarks, the $\Xi_{cc}\bar{D}^*$ and $\Xi_{cc}B^*$ states with $I(J^P)=0(1/2^-)$.

This information of these predicted triple-charm molecular pentaquarks and their partners may stimulate further interest in searching for them in near future.
In the past 14 years, abundant observations of charmonium-like $XYZ$ states has made the study of multiquark states become a hot issue of hadron physics. With the running of accelerator facilities such as KEK and LHCb, we have good reasons to believe that experimentalist will bring us more surprises. We also expect that these prediction of triple-charm pentaquarks come true. Obviously, more theoretical and experimental effort are needed.

\section*{ACKNOWLEDGMENTS}

This project is partly supported by the National Natural Science Foundation of China under Grant No. 11222547, No. 11175073, and No. 11647301 and the Fundamental Research Funds for the Central Universities. Rui Chen is supported by the China Scholarship Council. Xiang Liu is also supported in part by the National Program for Support of Top-notch Young Professionals. Atsushi Hosaka is supported by JSPS KAKENHI (the Grant-in-Aid for Scientific Research from Japan Society for the Promotion of Science (JSPS)) with Grant No. JP17K05441(C).

\end{document}